\documentclass[manuscript]{acmart}
\AtBeginDocument{%
  }

\usepackage{subcaption} 

\setcopyright{acmlicensed}
\copyrightyear{2018}
\acmYear{2018}
\acmDOI{XXXXXXX.XXXXXXX}
\acmConference[Conference acronym 'XX]{Make sure to enter the correct
  conference title from your rights confirmation email}{June 03--05,
  2018}{Woodstock, NY}
\acmISBN{978-1-4503-XXXX-X/2018/06}

\usepackage{xcolor}
\usepackage{colortbl} 

\definecolor{Butter}{RGB}{245, 216, 146}
\definecolor{Moss}{RGB}{192, 176, 91}
\definecolor{Palm}{RGB}{101, 118, 82}
\definecolor{Guava}{RGB}{242, 182, 163}
\definecolor{Sunset}{RGB}{232, 156, 115}
\definecolor{Sangria}{RGB}{227, 101, 89}
\definecolor{Seabreeze}{RGB}{229, 233, 235}
\definecolor{Lagoon}{RGB}{148, 190, 187}
\definecolor{Odyssey}{RGB}{35, 97, 126}

\begin{document}

\title[SocialWeave]{SocialWeave: Supporting Comment-supplemented Navigation of Video Content with Structured Engagement with Community Comments}

\author{Jingxian Liao}
\email{jxliao@ucdavis.edu}
\orcid{0000-0002-2474-4851}
\affiliation{
	\institution{University of California, Davis}
	\city{Davis}
	\state{CA}
	\postcode{95616}
	\country{USA} 
}

\author{Yi-Ting Hung}
\email{nctucs.cs08@nycu.edu.tw}
\affiliation{
	\institution{National Yang Ming Chiao Tung University}
	\city{Hsinchu}
	\country{Taiwan}
}

\author{Yi-Hong Lin}
\email{lin860707.eed04@nctu.edu.tw}
\affiliation{
	\institution{National Yang Ming Chiao Tung University}
	\city{Hsinchu}
	\country{Taiwan}
}

\author{Wen-Chieh Lin}
\email{wclin@cs.nycu.edu.tw}
\orcid{0000-0002-9704-5373}
\affiliation{
	\institution{National Yang Ming Chiao Tung University}
	\city{Hsinchu}
	\country{Taiwan}
    }
 
\author{Hao-Chuan Wang}
\email{hciwang@ucdavis.edu}
\orcid{0000-0003-3295-5111}
\affiliation{
	\institution{University of California, Davis}
	\city{Davis}
	\state{CA}
	\postcode{95616}
	\country{USA} 
}

\renewcommand{\shortauthors}{Liao et al.}

\begin{abstract}
People's navigation over online videos often leaves socially accessible cues, such as reactions, comment snippets. Drawing on Information Foraging Theory and prior work, these traces could act as "social scent" for cross-video navigation, yet they sit below the fold and accumulate beyond what people can leverage. This raises the need to understand how these ad hoc comments actually influence navigational behavior if available and how to make them accessible and actionable. We present SocialWeave, a video navigation interface that blends social comments with video-content-based guidance as navigation scaffold. In a study using YouTube data, we compared SocialWeave with content-based guidance only and a video playlist. We found that comments opened a parallel consultation channel, as participants consulted them at routing moments for verification and new directions. Our study reveals behavioral and stance factors impacting human interaction with algorithm-distilled comment cues, with design implications on AI-augmented search interfaces from diverse channels. 
\end{abstract}



\received{20 February 2007}
\received[revised]{12 March 2009}
\received[accepted]{5 June 2009}

\maketitle

\section{Introduction}


Modern video ecosystems now provide substantial support for information seeking and navigation. Among diverse channels, audience traces as \textbf{designed social cues} (e.g., view counts and even selective comment snippets in previews) also appear in the search results to help people quickly triage the videos in the video pools. 
They may strengthen a sense of social presence and serve as navigational guidance, following what Social Information Foraging Theory (SIFT) ~\cite{pirolli2009elementary} suggests. 
As people seek information by following information scents, users interpret cues (e.g., headings, tags, and comments) to estimate how promising candidate sources are for their goals. 
Lightweight social traces such as comments and annotations may guide users toward high-relevant sources and reduce search effort~\cite{nelson2009little}. 
When the comments are attached to single videos, time-anchored commenting (i.e, danmaku) can also help learners jump to segments based on others’ notes and popularity signals \cite{lee2015using}. 

Currently, audience comments are usually simply displayed and ordered with algorithmic prioritization, such as popularity, on video platforms \cite{ouyang2026commsense}. To leverage social comments at scale, researchers have turned to computational methods in domains such as music, public opinion sharing, and reviews. Sentiment classification and topic modeling \cite{chen2023amplifying} help aggregate content and reveal affective patterns and thematic threads across music comment pools. Recent LLM-based methods can also readily condense comment pools into summaries and themes \cite{liu2024selenite, chen2025commentscope}. However, these outlets arrive as standalone summaries, offering readers little structure to identify patterns of recurring topics, compare them, or return to the original content behind them.

Another line of relevant work attempts to add structure through visualization for browsing, interpretation, and reflection. For example, in order to help readers engage with comment content, the sets of comments are represented as maps of opinions and summaries \cite{faridani2010opinion}, visualized the sentiment-driven timelines \cite{sung2017exploring, chen2023amplifying}, and presented inline annotations anchored to the source text \cite{chen2025commentscope}.  Yet, most existing technical approaches to comment-driven navigation support have been confined to single-source settings \cite{lee2015using, sung2017exploring, ouyang2026commsense}. 
Across a video collection, a comment theme may relate to many videos at once, and interpreting or acting on it requires relating it back to the sources it reflects. Cross-video navigation requires both scaffolds: \textit{distilling large-scale comments into digestible overviews}, and \textit{keeping those overviews anchored to the videos and comments they arise from.}

However, such distillations at best serve as standalone summaries for analytical purposes, separated from the comments they condense and the videos those comments respond to; meanwhile, as comments' meanings are tightly coupled to the source content, cues separated from their sources are difficult to interpret and act on. Cross-video navigation thus calls for overviews that stay linked both downward to the underlying comments and outward to the videos they draw from. Although LLM-based summarizers can readily condense comment pools of massive size, the overall summary is not an overview useful for navigation purposes. A distilled text alone offers no support for comparing the community's reactions across videos, descending from recurring themes to aggregated voices. Second, as comments' meanings are tightly coupled to their source videos, pointing to a specific moment, they are difficult to interpret, let alone act on, when aggregated in isolation from the content itself. Addressing these challenges calls not for a well-performed summarizer but for a platform that couples comment aggregation with the source videos and renders the result navigable.

 \begin{figure*}
    \includegraphics[width=0.9\textwidth]{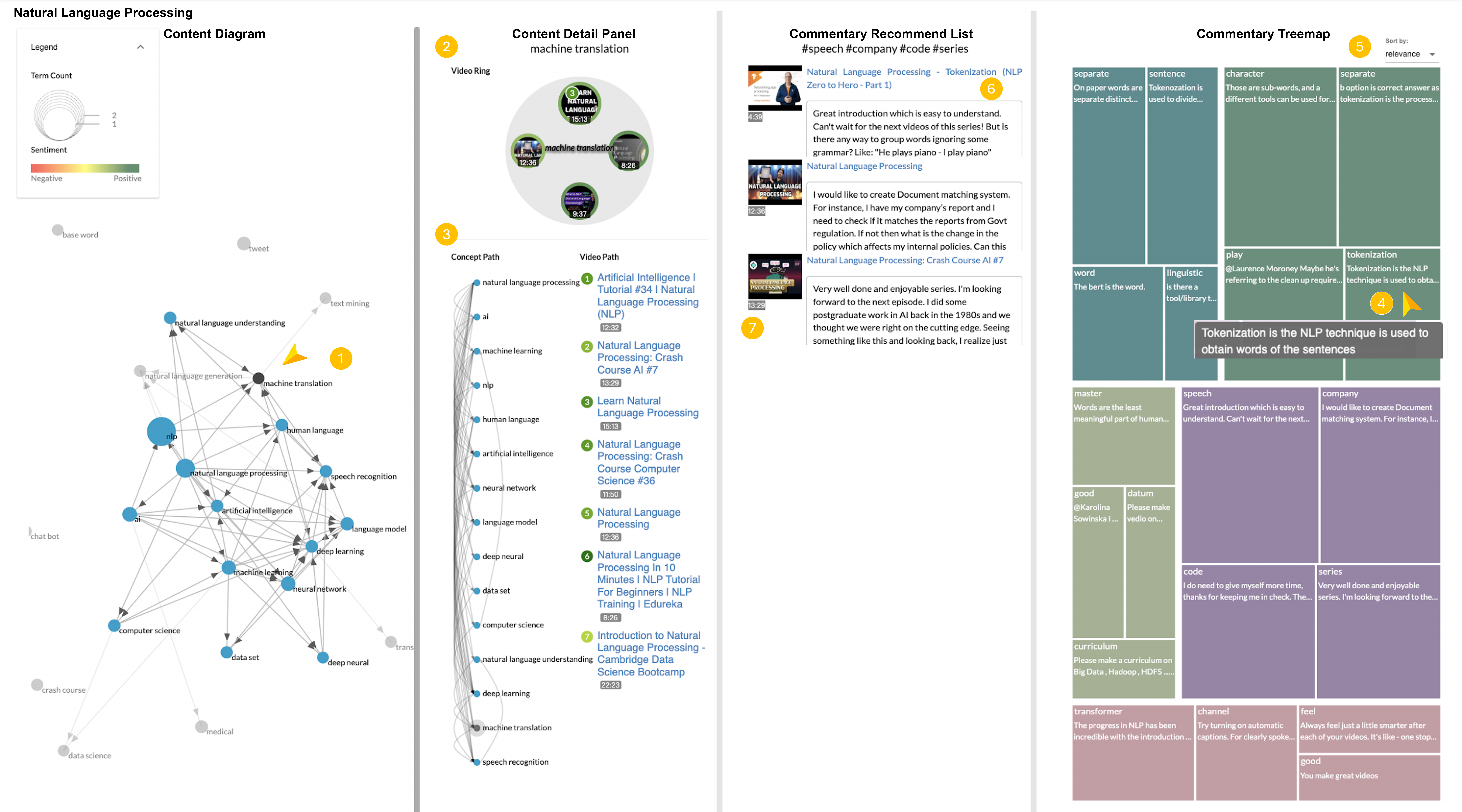}
    \caption{SocialWeave interface. Content Diagram: structural overview of video content where nodes represent top concepts and edges denote directional relationships (\textcircled{1}). Content Detail Panel: video recommendations for a selected concept, shown as a circular arrangement of videos and a sequential pathway from foundational concepts to the concept of interest (\textcircled{2}-\textcircled{3}). Commentary Treemap: structural overview of comments organized by theme, with representative comment snippets. Users can select theme nodes and adjust keyword ranking criteria (\textcircled{4}-\textcircled{5}). Commentary Recommend List: curated set of videos and representative comments for a selected theme (\textcircled{6}-\textcircled{7}). The four views are connected via the underlying videos. Hovering the concepts or comments will highlight their interconnected nodes and comment themes based on the underlying videos. }
    \label{figure:twopanels}
\end{figure*}


In this work, we aim to address the gap by designing a comment-guided video navigation interface, \textit{SocialWeave}. 
SocialWeave responds to the representational bottleneck in using comments for cross-video triage by pairing two linked landscape-level overviews: a content-derived overview that summarizes topical coverage and its relations, and a comment-derived overview that aggregates audience reactions into scannable themes that connect to the content overview. 
It builds upon the prior work that organize representations extracted from video content as the primary mediation layer, rendering cross-video structure as semantically linked sequences \cite{zhao2018flexible, schwab2016booc}. Users may seamlessly browse between both structural overviews and corresponding detailed views for video selection. Figure~\ref{figure:twopanels} presents a screenshot of SocialWeave, showcasing the topic ``Natural Language Processing,'' on a YouTube video set.



We conducted a lab study to understand how social comments impact video navigation routes, using real-world YouTube videos and comments. With 60 participants performing a learning-oriented navigation task, we compared SocialWeave with two baselines that represent common alternatives in practice and research: \textit{Content-centric Prototype}, ConceptGuide \cite{tang2021conceptguide}, which provides a content-based structured overview without community comments, and \textit{Default List}, a deterministic ranked playlist generated by YouTube recommenders with the viewer's profile disabled. Our findings revealed the complementary role of comments as new consultation channels for routing during the video navigation process, and also discussed the diverse factors that moderate the adoption of comment cues for navigation.

\section{Background and Related Work}

\subsection{Social Cues Around Videos}\label{sec:SocialCues}

Video platforms afford diverse social interaction channels (e.g., commenting, live chat, thumb-ups), enabling the audience to be exposed to a rich layer of textual and non-textual social signals that accompany viewing experiences. 
These interactions could occur both asynchronously and synchronously \cite{hrastinski2008asynchronous}. For live video streaming, real-time discussions between the audience or with the host improve social presence and encourage more real-time interactions \cite{lu2018streamwiki, fang2022understanding}. On general-purpose video platforms, people access recorded videos and leave and read comments displayed under the videos or time-anchored comments \cite{hrastinski2008asynchronous, hrastinski2009theory}.  Time-anchored comments, specifically, reveal the salient segments of videos from previous users, helping users to identify the highlights of videos and creating a pseudo-synchronous effect \cite{sung2017exploring, wu2019danmaku}.  
Tanprasert et al. also found that witnessing vicarious dialogues between peers and video owners alone can augment concentration and overall satisfaction \cite{tanprasert2023scripted}. 

Besides natural comments, social annotations and other interactions also play a role on social connections and attention guidance \cite{fang2022understanding,ko2022capturing}. For instance, social annotation of video transcripts may reduce the distraction in collaborative learning \cite{fang2022understanding}. 
Taken together, community contributions can facilitate attention guidance and navigation support. Building on prior work, this study explores the affordance of community comments for cross-video navigation. 

While prior work highlights the potential benefits of social comments, it does not eliminate the practical challenges to support cross-video navigation. Compared to creator-produced content, which provides a more information-dense signals, comments are at a different granularity. They surface situated audience reactions (e.g., confusions, disagreements, and takeaways) that can be valuable yet are distributed \cite{dutceac2022cueing, dubovi2020empirical}. Interpreting these reactions for navigation often requires contextual grounding in the video, sharing motivations, or domain knowledge, which can be difficult for novices \cite{bawden1999perspectives}. This mismatch is further compounded by the scale and low information density of comment spaces in online communities. On YouTube, videos receive highly uneven volumes of comments, and comments are typically presented below the fold, shaping their visibility and likelihood of being consulted \cite{schultes2013leave}. Moreover, comment sections also include discourse that is only loosely related to the video’s informational content (e.g., memes or self-promotion), increasing the effort required to locate navigation-relevant cues \cite{dubovi2020empirical}. Therefore, we assembled a customized computational pipeline to curate topic-relevant comments and organize them into scannable themes that connected back to candidate videos. 

\subsection{Analyzing and Visualizing Community Comments}\label{sec:CommentVis}

Prior work has developed a range of computational methods to distill large comment pools into digestible forms. Without the detailed content, the summary can be operated on affects, using sentiment and polarity classification to characterize the registers of comments~\cite{chen2023amplifying}. Content distinctions from topic modeling and clustering group scattered comments into a small set of recurring threads \cite{blei2003latent}. Where conversation itself carries structure, structure can be preserved rather than flattened as in recursive thread summarization, which can help readers descend into topics with chosen levels of detail \cite{zhang2017wikum}. LLMs have since relaxed the constraint of a pre-fixed summary target, making it possible to organize a group of comments around users' immediate information needs. For example, Selenite elicits comprehensive option-criteria overviews for online sensemaking, using GPT as a knowledge retriever \cite{liu2024selenite}. CommentScope applies LLMs to categorize comments into pragmatic types \cite{chen2025commentscope}. As the idea of comment summarization is widely shared, it becomes more flexible regarding how the distilled summary could be represented and organized for users who need to act on it across different scenarios.

Visualization research has charted most of the representations discussed above, primarily for a single comment pool. Encodings differ mainly in what they anchor the comments to. When comments carry a reference to the source content, such as a video timestamp or a paragraph, they are presented along that structure, sometimes stacked so as to expose where attention concentrates and how it shifts \cite{sung2017exploring}.  Absent such a reference, the anchor becomes the pool's own structure based on their semantic content or other data: comments are usually grouped by content similarity into topics and summaries, then projected into a landscape that readers traverse or arranged at varying levels of granularity \cite{faridani2010opinion, chen2025commentscope, wu2025comviewer}; sentiment alone has accumulated a substantial repertoire of such presentations \cite{kucher2018state}. 
Further strategies combine both views: pairing an organized overview with a display that returns comments to the content it addresses \cite{wu2025comviewer, chen2025commentscope}. 
These designs draw on the long-established idioms as in \cite{shneiderman1996eyes}: ``Overview first, zoom and filter, then details-on-demand''. Our interface design draws on the same vocabulary, and this work focuses on the abstraction they carry and the coordination that links it outward.

Beneath these visualization choices and analytical approach of comments, these studies are distinct in their purposes. In one group, comments themselves are the main foci: instructors analyzing time-anchored learner feedback \cite{sung2017exploring}, journalists interpreting viewer response \cite{brannon2024audienceview}, readers forming bias-aware opinions from a discussion \cite{ouyang2026commsense}, or shoppers weighing condensed reviews to choose between options \cite{yatani2011review}. In the other, they accompany engagement with source content, embedded at the passages of a long text they address \cite{chen2025commentscope}. This difference shapes which components each system aligns. When the pool is the focus, the summary links back to the comments it was derived from. When comments accompany other engagement, each comment is tied to both its summary and the source content it addresses. Cross-video navigation introduces a third relation. As the overview
summarizes a pool of pools, one comment theme is not tied to any single source, and aligning it means relating that theme to the several videos whose comments share similar content. The alignment thus groups candidate sources that share audience reactions,
supporting selection among sources. SocialWeave implements this alignment by anchoring each theme to the videos it draws from, so that an audience impression becomes an entry point to candidates.

\subsection{Navigation Interfaces for Video Collections}

To facilitate video exploration and browsing, existing research has pursued two primary avenues. The first focuses on aiding navigation within one single video with different cues, employing advanced algorithms and pipelines \cite{matejka2013swifter, liu2018conceptscape}. For example, user-generated comments and upvotes are used to build the summary of videos and time anchors within videos to help learners grasp the highlights of video content \cite{lu2018streamwiki, lee2015using, sung2017exploring}. 
Collective interactions with videos, such as replays and user progress and struggles, are encoded into the video timeline to help users plan ahead, guide pauses, and provide feedback during the process \cite{yang2022softvideo, pongnumkul2011pause}. 
For long videos, researchers have noted the difficulty of navigating through different sections of the video. Multiple systems have used AI algorithms to extract chapters from narration, images, and audio, some with editors in the loop, to generate clear video segmentation \cite{matejka2013swifter, fraser2020temporal}. More recent works have also begun to support semantic search by identifying visual and textual objects in videos using LLMs \cite{wang2024surgment}. 
Although these studies revealed the benefits of social interactions for content navigation, many studies collected comments from their participants exclusively for their studies, different from natural comments left on YouTube. Instead, our work grounds cross-video navigation in real-world comments, and in the perceptions users already bring to them, to understand when such social cues help and where they fall short. 

The second trajectory delves into navigation and visualization support across multiple videos, which is the focus of our work. They mainly aim to identify the connections between videos and provide interactive visualizations to help users navigate the video pool. For instance, MOOCex recommends videos via sequential relationships between videos and video topics, and further visualizes the semantic relationship of related videos on a 2D space with related videos in a Voronoi-alike diagram \cite{zhao2018flexible}. Surch~\cite{kim2023surch} displays a semantic graph of operation procedures and enables step-level video comparison. Content Flow Bar provides ``snippets'' of videos in a display with menus of keywords and definitions \cite{perez2022watch}. 
With generative AI, novel approaches emerge to restructure videos into snippet-based interfaces that remix segments across multiple videos and recompose narration flows, shifting video watching toward more task-oriented, focused experiences \cite{yang2025videomix, kim2023surch}. 
Most of these systems leverage video content and metadata; few studies have explored integrating audience-generated comments, especially from general-purpose video platforms. The lack of such opportunities can limit the social cues from the audience. Our work synthesizes ideas from these systems to mediate video navigation with overviews and proposes scannable overviews from social comments as a supplementary consultation channel. 

\subsection{Social Information Foraging Theory}

Social Information Foraging Theory (SIFT) may help understand how individuals seek, evaluate, and consume information in complex environments from a conceptual perspective, especially for social signals \cite{pirolli1999information, pirolli2009elementary}.  They observed that when seeking information, people behave like predators hunting prey. When navigating search results, they repeatedly evaluate which information sources (patches) to explore. People rely on information cues (e.g., headlines, keywords, links, or visuals) to assess the potential value of a patch. These cues contribute to what is known as \textit{information scent}, the perceived relevance or usefulness of a source based on cues for specific goals. A weak scent may prompt users to skip the patch and switch to a different source. 
Nelson et al. observed that individuals often forage social signals, such as annotations, comments, query histories, in sensemaking and other knowledge tasks \cite{pirolli2009elementary,nelson2009little}. 
Real-world applications are across massive platforms and communities \cite{wu2023transparentizing, chen2023amplifying}. 
These communities vary in size, closeness of connections, and structure. 
Our work focuses on social cues from weakly connected communities on video platforms: SocialWeave is designed to distill high information scent from comment content.

\section{SocialWeave Design}






In this section, we introduce three design rationales and developments of our comment-guided interface, SocialWeave. 
We grounded the design of SocialWeave based on the practical challenges of surfacing community comments as navigational cues (Section~\ref{sec:SocialCues}), prior work on comment visualization (Section~\ref{sec:CommentVis}), and Information Foraging Theory ~\cite{pirolli2009elementary}. 

\textbf{Distill large-scale, scattered community comments into a structured
overview.} Across a video set, user-generated comments accumulate at a scale that is impractical to read through~\cite{schultes2013leave}. Extending comment navigation beyond single-source settings, therefore, calls for aggregating these dispersed reactions into a digestible overview to help people extract informational content relevant to their temporal navigation needs ~\cite{pirolli2009elementary}.  
\begin{itemize}
  \item[\textbf{R1}] 
  The system should aggregate dispersed social comments
  across the video set into representative themes, and organize those themes
  with visual hierarchy and structure so that audience interests can be read at a glance rather than sequentially~\cite{shneiderman1992treemap,
  todorovic2008gestalt}.
\end{itemize}

\textbf{Weave social comment content together with video content to guide
cross-video selection.} Unlike video content, which is largely self-explanatory, comments derive their meaning from the source content they respond to~\cite{dubovi2020empirical} and are therefore difficult to act upon in isolation. Prior systems accommodate this dependency by aligning comments either with the summaries distilled from them or with the specific passages they respond to, in both cases keeping the source within reach as users read (Section~\ref{sec:CommentVis}). Cross-video navigation support requires the same traceability, yet it must serve a second end: during video exploration, the link from a theme back to its sources doubles as the means of reaching a candidate, supporting interpretation and selection at once. Traceability alone is not sufficient. Users pursue whichever cue proves salient at a given moment, so both channels must remain legible and mutually linked at the point of decision, extending the two-layer overview design of prior content-centric systems~\cite{tang2021conceptguide, kim2023surch}.
\begin{itemize}
  \item[\textbf{R2}] The system should anchor each comment theme to the videos and content it refers to, preserving access to the source comment and video so users can interpret the cue in context.
  \item[\textbf{R3}] The system should surface content and comment content together at the preview stage, supporting more informed and socially grounded selection.
\end{itemize}

Building on the ``overview-first, details-on-demand'' paradigm proven in prior content-centric navigation systems~\cite{tang2021conceptguide,kim2023surch}, we also design SocialWeave as two parallel content and comment overviews and detailed views, so users can forage compact overviews drill into specifics only when a cue looks promising, keeping the interface digestible rather than overwhelming \cite{shneiderman1996eyes}.

\subsection{Example Usage Scenario}

We first illustrate the experience of SocialWeave through a usage scenario, then detail each view and its interactions.

\textit{Leo, a self-directed learner, wants to understand Natural Language
Processing (NLP) on YouTube, which returns dozens of videos. Instead, he uses  SocialWeave to help him navigate. He begins with the Content Diagram, dragging to get an overview and noticing that ``language model'' and ``transformer'' sit at the center while ``machine translation'' lies toward the edge. Clicking ``machine translation'' opens the Content Detail Panel, whose video ring shows several video candidates with thumbnails, and a video path that sequences him from ``word embedding'' up to his target. Then Leo plans to watch the first video in the panel.}

\textit{Curious how previous learners actually receive these videos, Leo turns to the Commentary Treemap. Scanning its themes, he spots a recurring ``crash course'' keyword as he usually watches crash course videos on YouTube. He clicks on this theme, and the Commentary Recommend List returns a short set of videos with comments like ``Can't wait for the next videos of this series!'' Hovering the keyword lights up another video that connects with this audience's enthusiasm and also informative based on SocialWeave's highlights. Then he switches to watch this video through.}

\subsection{Visualization and Interaction Design} \label{sec:InterfaceDesign}

\subsubsection{Commentary Treemap}
On the far right of Figure~\ref{figure:twopanels}, \colorbox{Odyssey!50}{Commentary Treemap}  aggregates dispersed comments across the video set into scannable themes, surfacing the topics that audiences repeatedly raise (\textit{R1}). Comment clusters and their representative keywords are rendered in a squarified treemap layout (Figure~\ref{figure:twopanels} \textcircled{4}). The colors distinguish themes and rectangle size and position encode relative importance, arranging themes and keywords hierarchically for at-a-glance readability (\textit{R1}). Each keyword is paired with an exemplary comment snippet for understanding. Users can switch the keyword ranking criterion (Figure~\ref{figure:twopanels} \textcircled{5}) between frequency and semantic closeness to the video content to match their interest. 

SocialWeave enables multiple interactions to support users understand the details of the comment summary and connect with other views: hovering a keyword displays the exemplary comment snippet of the keyword. and also highlights the concepts mentioned in that comment's source video back on the Content Diagram, keeping the two content channels legible and linked (\textit{R2}). When clicking a keyword block as selecting, users activate the \colorbox{Lagoon!50}{Commentary Recommend List}.

\subsubsection{Commentary Recommend List}
\colorbox{Lagoon!50}{Commentary Recommend List} helps users focus on a compactly curated, comment-driven set of videos  whose audiences highly engage with the selected theme. Each block (\textcircled{6}) is accompanied by resonant comments, video title and thumbnail, on the second right of Figure~\ref{figure:twopanels}. 

Beyond the basic metadata, SocialWeave also checks the transcripts and all comments of these focused videos. Hovering a video accentuates its concepts mentioned on the \colorbox{Sunset!50}{Content Diagram} and its comment keywords on the \colorbox{Odyssey!50}{Commentary Treemap} (Figure~\ref{figure:hover} \textcircled{7}), letting users have a preview from an audience-first entry point while still grounding each candidate in the underlying content (\textit{R3}).

\subsubsection{Content Diagram and Content Detail Panel}
\colorbox{Sunset!50}{Content Diagram} and its \colorbox{Sangria!50}{Content Detail Panel}  visualizes the semantic structure of the video set as a directed graph of concepts, adopting the designs paradigm of ConceptGuide~\cite{tang2021conceptguide}. Basic and core concepts sit at the center of the canvas, encircled by more advanced or peripheral ones, and node size encodes a concept's relative frequency across the videos. Directed links indicate progression from basic to advanced concepts with a force-directed algorithm. Selecting a concept node opens the \colorbox{Sangria!50}{Content Detail Panel}, which surfaces a group of recommended videos and viewing sequences for that concept across three views: a video ring, a concept path, and a video path. In the ring, candidate videos are arranged clockwise as thumbnails, and the color of each outer ring encodes comment sentiment: the darker the green, the more positive the reception. The concept path traces the selected concept back to its prerequisite concepts, and the video path turns that trace into a followable sequence from the foundational to the target concept.

\begin{figure}
        \centering
        \includegraphics[width=0.7\textwidth]{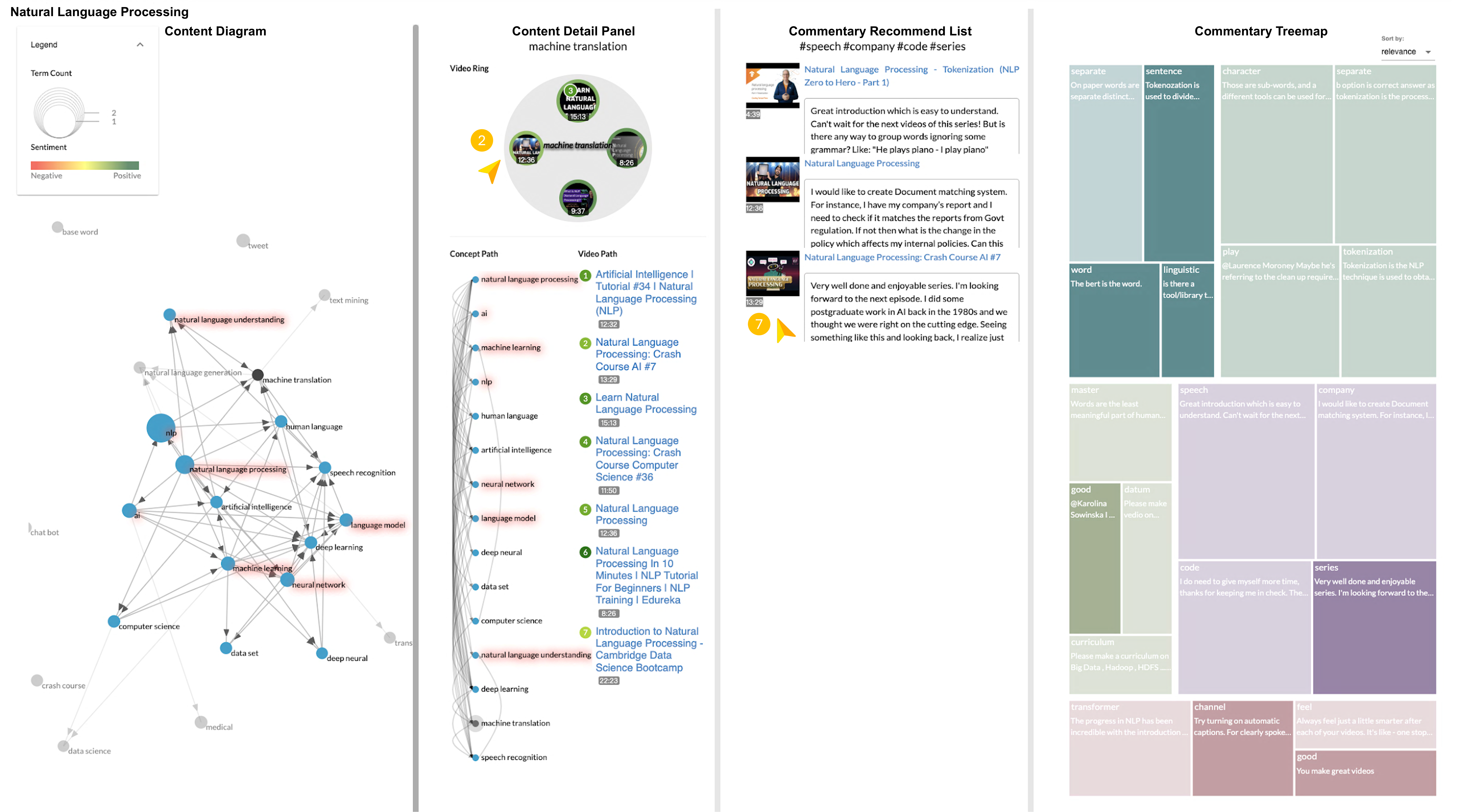}
        \caption{Highlights when hovering over a video on Content Detail Panel or Commentary Recommend List}
        \label{figure:hover}
\end{figure}

Critically, the four views are all connected via the underlying videos. Hovering any concept node highlights its interconnected nodes and links based on the underlying videos. Hovering or selecting a concept identifies the leading video for that concept and highlights its comment keywords in the treemap while fading others, giving users an immediate sense of how content and community discussion align (\textit{R2}). Also, in \colorbox{Sangria!50}{Content Detail Panel}, hovering any video previews its content and comments in place as in \colorbox{Lagoon!50}{Commentary Recommend List}: The concepts it covers and its comment keywords will be highlighted on two summary overviews (Figure~\ref{figure:hover}), so users can weigh content coverage and audience reaction together at the point of decision (\textit{R3}).

Once a user clicks a video in either middle panels, SocialWeave opens the
video player view: a video player, \colorbox{Sunset!50}{Content Diagram} for concept orientation, and a re-ranked short list of the comments that drew the user
there as the commentary treemap may not be representative for the specific video. They support a smooth transition from overview-level triage to focused viewing, and facilitate the users during the video watching.

\subsection{Computational Pipeline}\label{sec:SystemConstruction}

\begin{figure}
\includegraphics[width=1.0\textwidth]{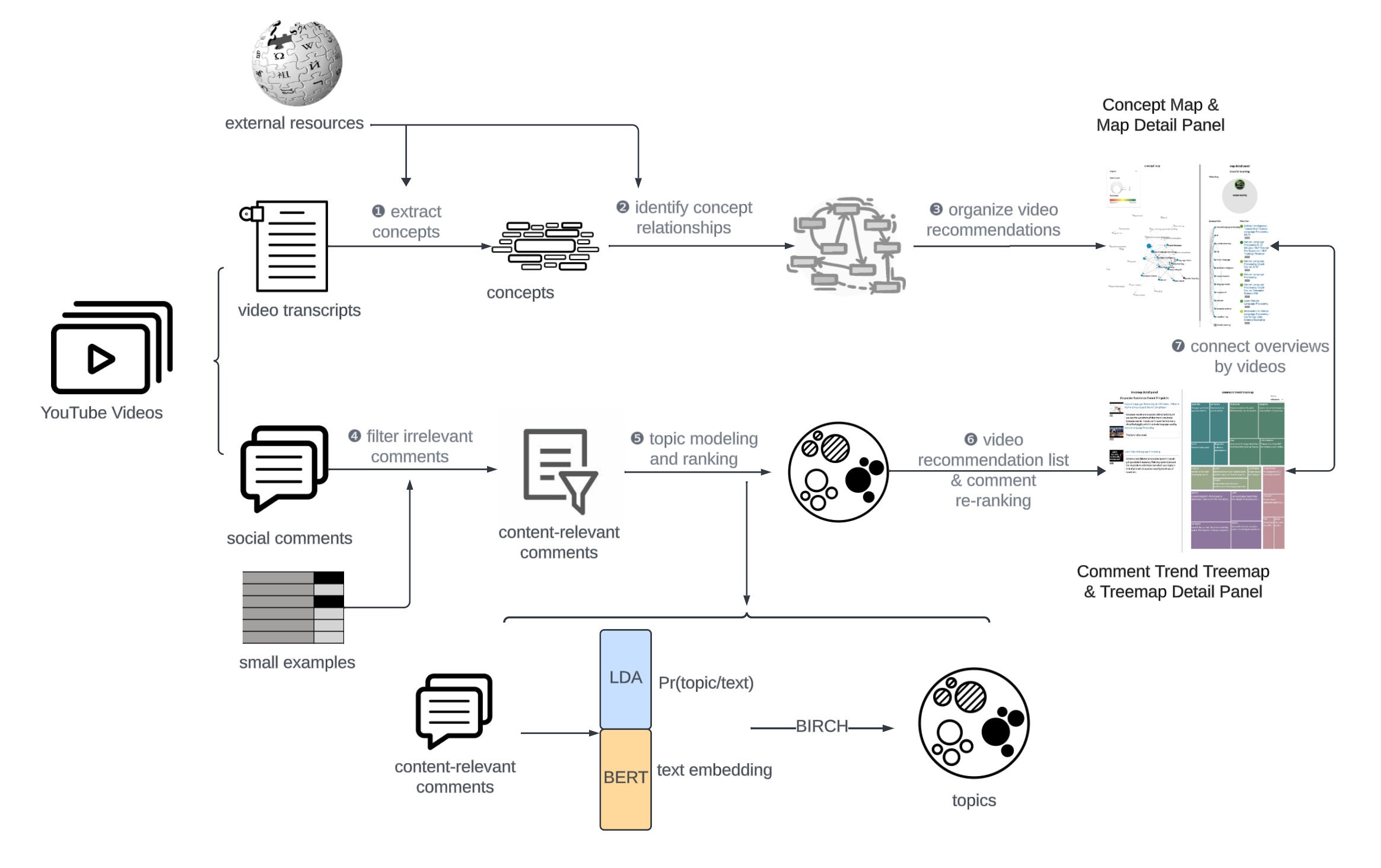}
\vspace{-2em}
    \caption{SocialWeave Computational Pipeline }
    \Description{}
    \label{figure:pipeline}
\end{figure}

SocialWeave is powered by a computational pipeline that transforms raw videos and their audience comments into the two linked structural overviews. There are four key steps to generate the underlying data as shown in Figure~\ref{figure:pipeline}: assembling a video corpus, deriving the \textit{Commentary Treemap} from comments, deriving the \textit{Content Diagram} from transcripts, and ranking videos and comments into the lists that link both overviews back to their sources. 

\textbf{Design decision.} We build the pipeline by adapting and fine-tuning language models rather than by prompting a general-purpose LLM or retrieving over the corpus with RAG. LLMs are now capable of capturing core concepts across heterogeneous sources, but the knowledge maps they produce lack coherent hierarchies and contextual grounding \cite{he2025great}. More importantly, SocialWeave's interaction designs require \textit{interpretable, local} links between every concept, theme, and keyword and its source videos and comments, which is difficult to guarantee with free-form generation. We therefore fine-tune language models and stack them with statistical and rule-based measures, calibrating the outputs through inspection until the resulting views held together and comprehensively represented the corpus. It helps to balance the performance between computational efficiency and manual controls \cite{devlin2019bert, sun2021ernie}.

\textbf{Implementation.} SocialWeave is a full-stack application: a Flask backend serving a frontend built in JavaScript, React, and D3. The computation pipeline runs with  Hugging Face \cite{HuggingFace} and PaddleNLP \cite{paddlenlp}.

\subsubsection{Data Assembly}

For a search query, SocialWeave pulls the top-ranked non-short videos from the YouTube Data API \cite{YoutubeAPI} together with their captions, comments, and metadata: titles, thumbnails, like counts, duration, descriptions, and channel information. Two rules then narrow the set based on our empirical observations. Videos need to carry English captions, whether auto-generated or uploaded, as the system mainly supports English now. Videos needs also run between four and thirty minutes, a window we set from the corpus deep dive. Short video commands enormous attention among younger viewers, but it trades depth for nano-content and edutainment \cite{khlaif2021using}, so we leave it out in our corpus. Filtering typically retains at least half of what the query returned. We also drop comments that are too brief or carry little information. 


\subsubsection{Commentary Treemap Formation}\label{sec:treemapconstruction}

SocialWeave first filters informative comments, clusters them into themes, and represents each theme with a compact set of keywords and exemplar comments, turning dispersed audience discourse into a scannable overview as shown above (\textit{R1}).

\textbf{Comment selection.} User comments are generated for diverse purposes and have a long-tail effect in terms of volume due to the popularity \cite{madden2013classification}, from personal sharing and entertainment to self-promotion, which introduces noise and inconsistency in their informational value for navigation. 
The pipeline first samples the top 100 or 40\% most-liked comments per video, balancing the comment spectrum between highly engaged videos and those with less audience interaction.
A customized comment classifier built on ERNIE   \cite{sun2021ernie} then remove off-topic content. The classifier adds a linear layer over the output layer, which we fine-tuned few-shot on comments coded as relevant/irrelevant based on \cite{madden2013classification}. 

\textbf{Comment topic modeling.} As comments are largely contextual to the videos, word co-occurrence methods alone can miss the connection semantically. In our initial exploration, methods such as LDA produced fragmented and unrepresentative clusters. Inspired by \cite{contextual_topic_identification}, the pipeline uses a hybrid clustering model that concatenates LDA topic and SciBERT sentence embeddings as features \cite{Beltagy2019SciBERT}. It then uses BIRCH clustering \cite{zhang1996birch} to select 4–6 themes based on coherence scores for visualization.

\textbf{Keyword and exemplar selection.} Each theme is represented by two complementary keyword sets, matching the two ranking criteria users can toggle in the interface (Figure~\ref{figure:twopanels} \textcircled{5}). The frequency-oriented set identifies common keywords via TF-IDF; the semantically-oriented set starts from the most frequently mentioned candidates and re-selects the ones with the highest sentence-level semantic similarity to the corresponding video transcripts, computed with SentenceTransformers \cite{reimers2019sentence}. To help users interpret each keyword, the pipeline also pairs it with a representative comment, which usually contains this word with the highest keyword–comment similarity. For example, "renewable energy" pairs with "The infrastructure to produce and run green renewables is also powered by conventional resources like coal."

\textbf{Video and comment linkage (\textit{R2}).} For each comment theme, the system recommends videos by measuring sentence-level semantic similarity between video transcripts and the theme's keywords using SciBERT embeddings, curating the top videos as the Commentary Recommend List. While watching, comments are re-ranked by a weighted score combining each comment's semantic similarity to the transcript and its like count. Because there is no ground truth for optimizing such a ranking, we adopted this pragmatic weighting to balance topical relevance with community endorsement.

\subsubsection{Content Diagram Formation} \label{sec:mapconstruction}

Content Diagram captures the key concepts from video transcripts and their association. To create it, we extracted concepts and then identified the relations between them, adjusting the pipeline used in Tang et al.'s ConceptGuide \cite{tang2021conceptguide}. We summarize the pipeline and our adjustments below. 

\textbf{Concept extraction.} In our early attempts, no single extraction technique proved robust across topics. So the pipeline stacks complementary sources of evidence. It applies KeyBERT in two rounds (over each transcript individually, and over the pooled transcripts) to surface keywords recurring across videos \cite{grootendorst2020keybert}. It also collects producer-supplied video tags as candidates and matches domain terms from external knowledge resources including DBpedia and Wikipedia against the transcripts \cite{wikipedia2026, auer2007dbpedia}, with Google's NLP API \cite{GoogleNLP} to ascertain the domain of the topic query. All candidates are then ranked by relevance importance scores, lexical length, frequency across multiple videos, and presence in video titles as in \cite{tang2021conceptguide}. The top 30 concepts are retained, a size we calibrated to balance visual clarity and information density.

\textbf{Relation mapping.} To structure the concepts into a graph, the system scores each concept pair (a, b) with four rule-based measures (Table \ref{tab:relations}), building on Pan et al. \cite{pan2017prerequisite}. They are semantic similarity, video and Wikipedia reference prerequisites, and complexity level distance. The pipeline combines them into a normalized prerequisite score whose sign gives the link direction; pairs above a threshold become directed links in the diagram. Detailed definitions and calculations are provided in Appendix. Finally, for each selected concept, the pipeline recommends a ring of videos ranked by concept frequency (with comment sentiment from TextBlob \cite{loria2018textblob} encoded on the thumbnails) and derives a viewing sequence by topologically sorting the prerequisite graph from basic to target concepts.

\begin{table}[th]
\centering
\begin{tabular}{ll}  
\toprule
\textbf{Metrics} & \textbf{Detail} \\  
\midrule
Semantic similarity (Ss) & Mean of cosine similarity of \( a \) and \( b \), co-occurrence in videos, and Wikipedia pages.\\  
Video reference prerequisite (Vrp)  & Relative positions of \( a \) and \( b \) in videos. See Appendix~\ref{appendix: comptutation} for calculations. \\  
Wikipedia reference prerequisite (Wrp) & Relative positions of \( a \) and \( b \) in Wikipedia. See Appendix~\ref{appendix: comptutation} for calculations. \\  
Complexity level distance (Cld) & \( avc(a) - avc(b) \), where \( avc(a) \) is the proportion of videos referencing \( a \).\\  
\bottomrule
\end{tabular}
\caption{Measures for the prerequisite relation between concepts $a$ and $b$}  
\label{tab:relations}  
\end{table}

\subsubsection{Pipeline Validation}

As no ground-truth concept structure or thematic partition exists for most video corpora, such as YouTube, we validated the pipeline against a few principles that the output should meet for the content and to support the SocialWeave design: representativeness, coherence, visualization legibility, and clarity. Especially, the outputs should reflect the video corpus and its audience rather than a few dominant videos and comments, channel-specific jargon, or promotional metadata. And comment themes should remain distinguishable from one another and interpretable; while the relations of concepts, which range from foundational to dependent content, also remain locally interpretable.

Three researchers had independently reviewed pipeline outputs against these four criteria, adjudicating by majority and returning to the source video or comment to resolve contested cases. Review proceeded over 15 topic corpora spanning mathematics, physics, computer science, social science, news, and several emerging topics like NFT. These criteria bound the rules introduced above, such as the size of groups and keywords, and the filtering of comments. The configuration was frozen before data collection and applied unchanged to the two corpora for the study. It suggests that the pipeline could support structuring  dispersed video content and community commentary into coherent, source-traceable structures for navigation. Also, the focus of this work lies primarily in the pipeline abstractions, the structural representations of content and social cues, and the interaction designs that connect them, which are independent of specific model choices. As model performances advance, we anticipate these design choices will remain valid with improvement in output quality. 

\section{User Study}

We conducted a user study to understand how users decide their navigation processes with the comment-derived summary and cues. We compared three video navigation interfaces: (1) SocialWeave, (2) ConceptGuide \cite{tang2021conceptguide}, which 
provides a content-based structured overview without comments, and (3) a video list from YouTube search results as a resource-controlled baseline. It follows the between-subject design. 

    

\subsection{Conditions}

The study comprises three conditions with variations in the availability of structural representational overviews. These three conditions consist of different abstractions and visualizations of information for navigation with the same video pool. 

\emph{SocialWeave}. \quad Participants use SocialWeave to explore videos, which include both video content and social comment overview, along with detailed navigation panels. 

\begin{figure}
    \begin{minipage}{.55\textwidth}
        \centering
        \includegraphics[width=\linewidth]{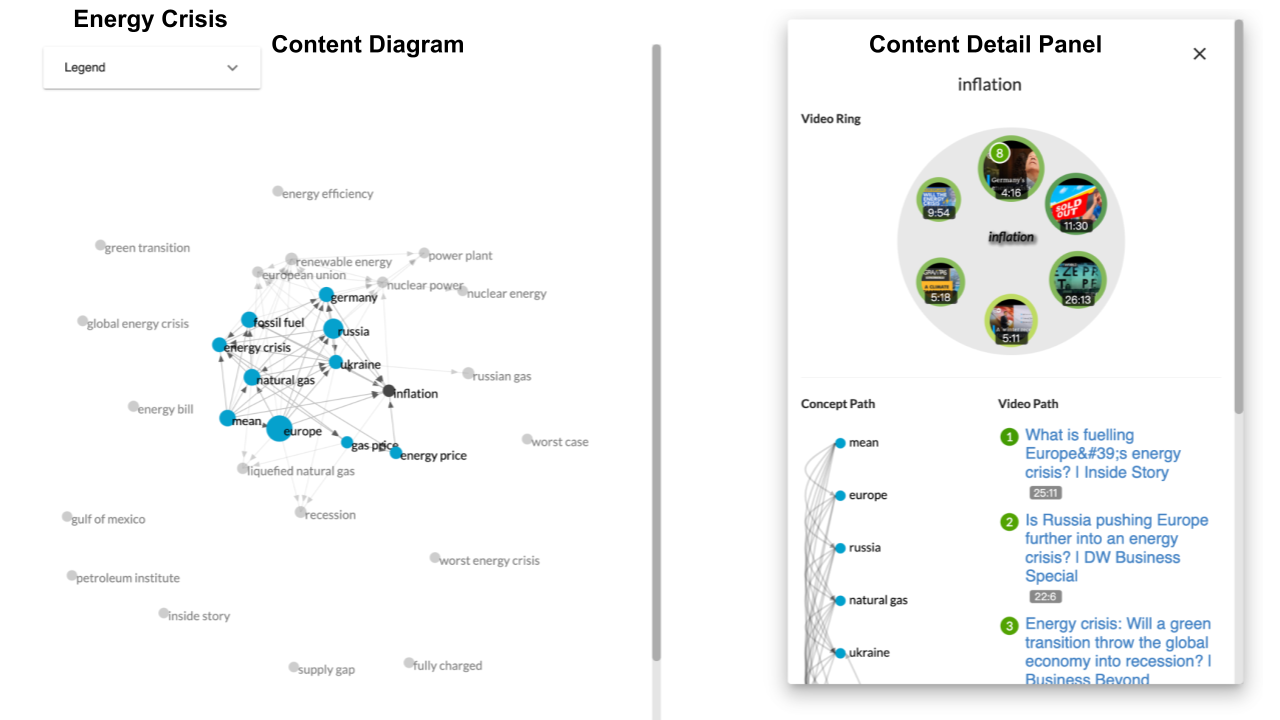}
        \caption{Content-centric Prototype \cite{tang2021conceptguide}}
        \label{figure:cgplot}
    \end{minipage}
    \begin{minipage}{.35\textwidth}
        \centering
        \includegraphics[width=\linewidth]{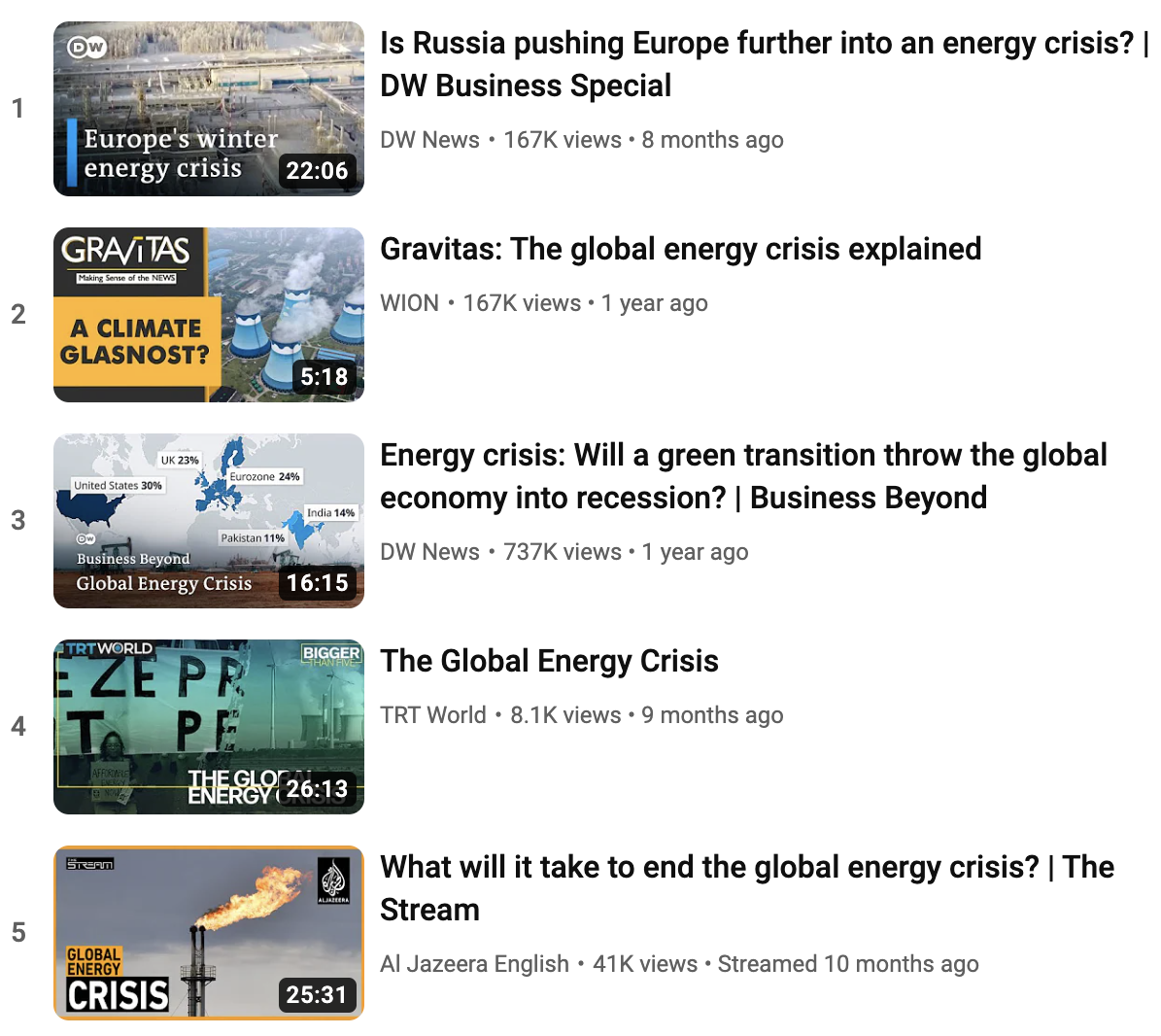}
        \caption{Default List}
        \label{figure:ytplot}
    \end{minipage}
\end{figure}

\emph{Content-centric Prototype}. \quad 
In the Content-centric condition, participants engage with an interface that presents structural video content cues only, introduced by Tang et al \cite{tang2021conceptguide},  
as shown in Figure \ref{figure:cgplot}. It serves as an example of research innovations that primarily utilize video content (e.g., text or images) and metadata.

\emph{Default List}. \quad In this condition, participants were presented with a video list only (shown in Figure \ref{figure:ytplot}), a baseline that approximates current navigation information on the YouTube search result page. 
The sequencing of these videos followed the default results generated by YouTube's recommendation algorithms. To generate this list for each specified topic, we utilized the YouTube Data API \cite{YoutubeAPI}. The videos were then arranged in the order they were returned by the API. Please note that the default list simulates the page after the user's initial keyword search. It's chosen to control the video scope across conditions. 

\subsection{Participants}

Through emails and school social media, we recruited 66 participants from two university campuses in different countries. The participants were required to be at least 18 years old, fluent in English, and have experience with video search and watching informative videos. They come from diverse backgrounds, including colleges and majors, video-watching frequency, and online learning frequency.
Among them, three were pilot participants, and three others were removed due to missing data or attention check failure during the study. In the end, there were 20 participants in each condition, aged 18-37 (mean=22.98, sd=3.90). Thirty-five of them self-identified as female, 23 as male, and the other 2 preferred not to disclose or be disclosed as non-binary. In terms of their video learning frequency, 58.3\% of participants watch online videos for learning one to multiple times a week, 16 participants learn through videos every day, and others learned via videos less often. As participants' disciplinary backgrounds differed, we screened them before the study, and no one self-reported expertise in the topic they were assigned to watch. 
We ran Kruskal-Wallis tests to compare numeric demographic metrics across conditions, as the values were not normally distributed. We found no significant differences between conditions in video learning frequency ($\chi^2(2)=0.02$, $p=0.98$) and familiarity of the topic ($\chi^2(2)=1.96$, $p=0.39$). We also confirmed no differdxxence in gender and education level across conditions via Fisher's test, given the size of the data (gender: $p=0.18$, education: $p=0.95$). The study lasted about 90 minutes, and participants were compensated with \$12 for their time. The study was approved by both campuses' ethics review boards.

\subsection{Study Protocol}

This study is a between-subjects experiment. Participants were also randomly assigned to learn one of two distinct topics in a random condition: Natural Language Processing (NLP) or the European energy crisis. The distribution did not differ significantly across conditions ($p=0.99$ in Fisher's test). Each participant was given 60 minutes to navigate and watch the videos, with a 27-inch monitor and a desktop.  
Participants first signed the consent form and completed a pre-study survey for demographics and prior learning experience. Participants in SocialWeave and \emph{Content-centric Prototype} conditions then spent 5 minutes watching a brief tutorial explaining the main functions of the prototypes. Before the experiment, participants also completed a hands-on practice session with another example topic, to increase their familiarity with the system.  
They were instructed to watch the videos thoroughly and at their own pace. They could take notes, change video player speed, etc., as they typically do. Upon completion, participants completed a survey that evaluated their video navigation experience and reflected on the processes, such as learning satisfaction, perceived guidance. 
We also conducted semi-structured interviews around the usage of the interface and the video navigation processes. Participants shared the factors that influenced their video navigation and selection. 

\subsection{Videos Materials}

Two video topics are chosen for the study: Natural Language Processing (NLP) and the 2022 energy crisis in Europe. Both topics attract broad interest from general YouTube audiences seeking information~\cite{googletrendec}, and encourage comments. We drew them from different domains so that our findings would not hinge on a single style of content or audience discourse.  NLP is a technical, concept-dense subject with a relatively stable body of knowledge, where comments tend to reflect questions or interesting points in the videos. The energy crisis, by contrast, is a news-driven topic spanning economic, environmental, and social concerns, where comments more often express opinions and personal interpretations. Including both topics helps ensure that our observations of navigation are from diverse content domains and discourse styles, rather than reflecting the idiosyncrasies of one topic. 


For the NLP topic, 17 videos were selected. The NLP corpus contains 17 lecture-style videos uploaded over the years, using slides, whiteboards, and occasional animation. They originate from industry leaders like Google and IBM, as well as reputable YouTube channels such as Crash Course. The Energy Crisis corpus included 24 videos. These videos primarily consist of news reports, interviews, and documentaries produced by mainstream media outlets. Videos in each topic vary in presentation style, content, and social popularity, as measured by view and comment counts, reflecting the diversity of videos available on general video platforms.

Before the study, we confirmed that the comment themes in SocialWeave were not already visible in the default comment list. We located, for each of the 41 videos, the earliest comment in which each of its themes appeared, ranked by like count, which closely approximates YouTube's default \emph{Top comments} ordering. Across all resulting (video, theme) pairs, a video's five highest-ranked comments covered only 16\% of the comment keywords appeared in the themes, its top ten 28\%, and its top twenty 41\%. Coverage was lower for the energy crisis videos (12\%, 20\%, 32\% for the top five, ten, and twenty) than for NLP (30\%, 50\%, 67\%), reflecting their larger comment pools. Therefore, a major portion of the themes surfaced from comments below the fold. 

\subsection{Data and Analysis}

\subsubsection{Video Navigation Logs}

To understand video navigation process, especially for SocialWeave, we analyzed participants’ video selection process, the channels they consulted, and video watching history, during the 60 minutes. We extracted three log-based measures of navigation process: whether participants changed their minds about which video to watch (\textit{video switch}), which panel prompted that change (\textit{switch trigger}), and how much of their consultation went to the comment channel over the course of the session (\textit{comment hover share}) for SocialWeave. Most of these measures apply only to Content-centric Prototype and SocialWeave, since the Default List offers no cue or candidate panels whose consultation could be logged. The measure also applies to Default List when it comes to video watching history.


Each participant's interaction log is segmented into \textit{decision episodes}: the stretch of activity between finishing one video and starting the next. 
The first episode for this participant is dropped because it conflates deliberation with initial orientation to the interface. 
In these decision episodes, \textbf{video switch} captures the occasions of one's mind mid-deliberation changing: the participant was already leaning toward one video, paused to consult some new information, and then went on to watch a different video after that consultation. As the hover actions are common, only clear signals that new information is consulted are counted, for example, a new theme clicked, or multiple hovers around different panels that are at least as much as they typically hovered the videos they looked at.  For every switch in SocialWeave we additionally coded its \textbf{trigger}: a switch is comment-triggered if the consultation was from the Commentary Treemap, or if the video finally chosen came from the Commentary Recommend List, and content-triggered otherwise. Switches in the Content-centric Prototype are content-triggered by construction. Finally, for each SocialWeave participant, we computed the hover count of different panels over time, and calculated the proportion by channels in successive 5-min windows. This proportion is referred to the \textbf{comment share} of the hovers: the number of hovers on the Commentary Treemap and Commentary Recommend List divided by the total hovers on both channels in that window, where the content channel comprises the Content Diagram and Content Detail Panel.
Moreover, we capture early abandonment via \textbf{video dropout} count: videos watched for less than 60 seconds or with a watch rate (time watched / video duration) below 10\%, thresholds chosen to capture cases where participants opened a video but abandoned it before engaging with substantive content.  
These metrics are aggregated at the participant level for analysis, so that people who watched more videos or had more actions do not weigh more heavily in the comparison. 
Together, these measures capture behavior at routing junctions and whether the video chosen were participants stayed with. 




\subsubsection{User Experiences}

We use surveys to understand how users explore videos, especially in Content-centric condition and SocialWeave. 
Participants also completed the NASA-TLX to assess their cognitive load during the task \cite{hart1988development}.  


\subsubsection{Analysis} 

To analyze the quantitative data, we primarily used mixed-effect ANOVA and Mann-Whitney U test. For ANOVA, the video topic and recruitment site were used as random variables. Participants' familiarity with the video topic, frequency of online video learning, and other demographic information, such as gender and education level, were used as covariate variables. We performed Tukey comparisons for post hoc comparisons. Mann-Whitney U test was used for the data that were not normally distributed, such as video switch. And the rank-biserial correlation is reported for effect size.



For the interview, we open-coded participants' interview responses on their use of comment-related cues and the reasons behind it, and how they used social cues to make navigation decisions \cite{strauss1998basics}. Two researchers independently performed initial coding, generating recurring codes from the transcripts. All authors then iteratively refined these codes into a shared codebook through discussion until consensus was reached, following a consensus-based qualitative analysis approach.

\section{Results}\label{sec:result}






\subsection{Navigation Outcomes}

We first report the videos that participants ultimately chose, and then dive deep into the navigation process and the channel consulted. 

Across conditions, participants ultimately accessed similar \textbf{knowledge coverage}, and the availability of SocialWeave changed which videos they committed to and how much redundant exploration they incurred. Participants watched 7.45 videos on average (sd=2.02) and these videos mentioned similar amounts of video concepts across the conditions (Default List: mean = 23.60, sd = 3.57; Content-centric Prototype: mean = 24.45, sd = 2.46; SocialWeave: mean = 24.15, sd = 3.30).  The amount of comment information they encountered was also comparable across conditions, in terms of the number of comment keywords (mean = 25.89, sd = 4.97). These coverage measures did not differ significantly across conditions, suggesting that all enabled participants to reach roughly the same breadth of content.

We also compared the \textbf{video dropout} count. There is a significant difference in video dropout due to the differences in structured overviews ($F(2,50)=12.91$, $p<0.01$, $\omega^2=0.31$). Participants using the Default List dropped many more videos (mean = 5.05, sd = 4.12) than those using the Content-centric Prototype (mean = 1.10, sd = 1.37; $z=4.61$, $p<0.01$) or SocialWeave (mean = 1.30, sd = 1.75; $z=4.03$, $p<0.01$).  It suggests that the semantic scaffold contributed to efficient navigation, reducing the need to “sample and abandon” videos in the player, and that the addition of Commentary Treemap preserved these benefits without complicating the navigation.
 


\subsection{Consultation at Routing Moments}

Having found comparable content coverage, we turn to what the comment channel changed about the decisions themselves. When looking at the video selection processes, participants using SocialWeave show switching videos more often than they had considered before watching them. In the SocialWeave and Content-centric Prototype, there are comparable numbers of video selection decision points after the first video watched: 6.75 in the Content-centric Prototype (sd = 2.12) and 7.45 in the SocialWeave (sd = 3.17) per participant. There are no significantly additional junctions due to the comment panels. 
However, participants using SocialWeave \textbf{switched} away from their front-runner video in a mean of 12.5\% of their decisions per person (sd = 0.11), and people in the Content-centric condition switched by 5.3\% (sd = 0.08). A Mann-Whitney U test shows this as a significant difference that participants made more video selection changes when the social channel is available ($U=275$, $p=.031$, rank-biserial $r=+0.38$).

Further comparison of \textbf{switch triggers} showed that the increase in switching and reconsideration was due to the comment channels. (As the switch count is not sufficient for the statistics test within the SocialWeave condition, we report the descriptive results here.) Of all the switches in SocialWeave, 5 participants followed content-derived panels, and 9 switched their videos after consulting the Commentary Treemap. In Content-centric Prototype, 7 of 20 have switched videos. Therefore, the additional rerouting was based on the comment-based panels rather than on the more intensive use of the content-based panels.

We complemented these behavioral measures with participants’ self-reported cognitive load. Kruskal-Wallis test is conducted to compare NASA-TLX scores across the three conditions and did not observe a statistically significant difference ($\chi^2=5.27$, $p=0.08$; Default List: mean = 3.77, sd = 0.65, Content-centric Prototype: mean = 3.37, sd = 0.90, SocialWeave: mean = 3.86, sd = 0.80).


The interviews indicate the considerations behind these consultation moments. Multiple participants described consulting the treemap at transitions, such as when deciding \textit{``whether to explore deeper or change another direction''} (P303). 
The comments sometimes function as verification of the video content. Three participants recalled they used comment volume and theme color to confirm a choice they had already provisionally made, for instance, to \textit{``choose well-received, commented videos''} (P315). P310 described how they cross-checked two sources: they returned to the Content Diagram when the comments diverged from a video's framing (\textit{``Some comments were not the same as the idea of the video''}), using the comments to raise a question about a video candidate. Two participants both identified one unfamiliar comment theme as a new topic that they might not have noticed from the video content panels (e.g., “\textit{saw the comment ‘green transition’ and selected a video there [Commentary Recommend List]}”, P312, P313). Participants consulted the comment summary and details to explore, confirm, and reselect candidate videos.

\subsection{Temporal Staging and Stance of Social Comments in Consultation}

\begin{figure}[htbp]
    \centering
    \begin{subfigure}[b]{0.8\textwidth}
        \centering
         \includegraphics[width=0.8\textwidth]{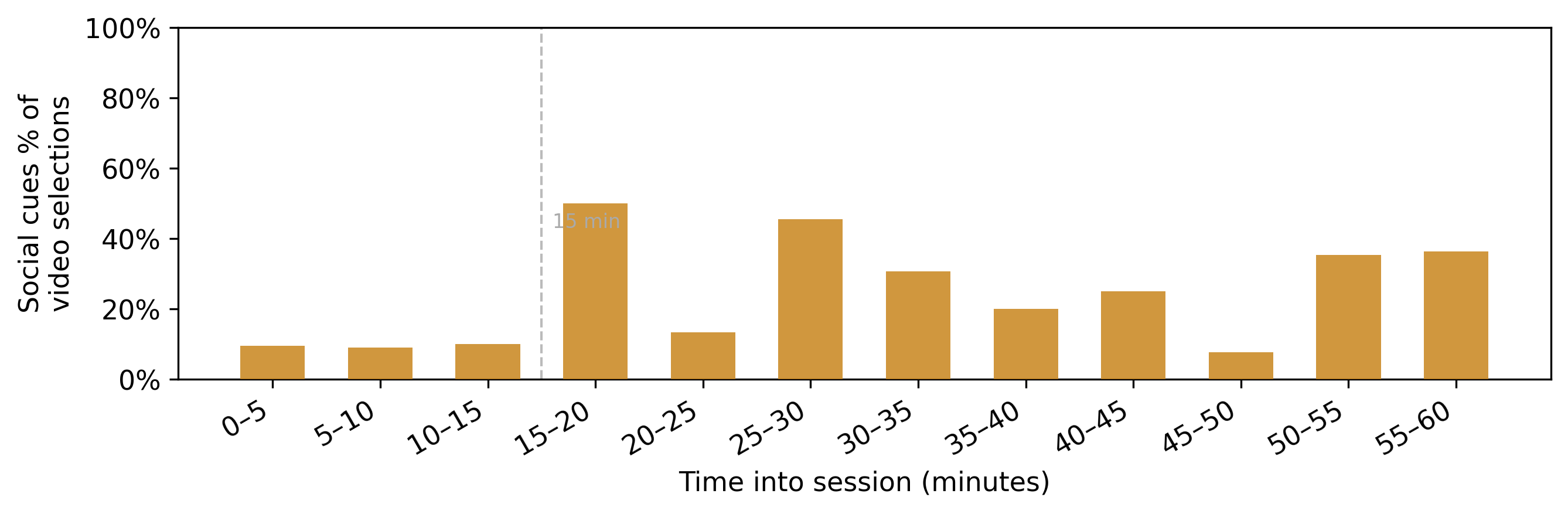}
        \caption{Share of video selections via social cue panels in SocialWeave.}
        \label{fig:treeclick}
    \end{subfigure}
    
    \vspace{1em} 
    
    \begin{subfigure}[b]{0.8\textwidth}
    \centering
         \includegraphics[width=0.8\textwidth]{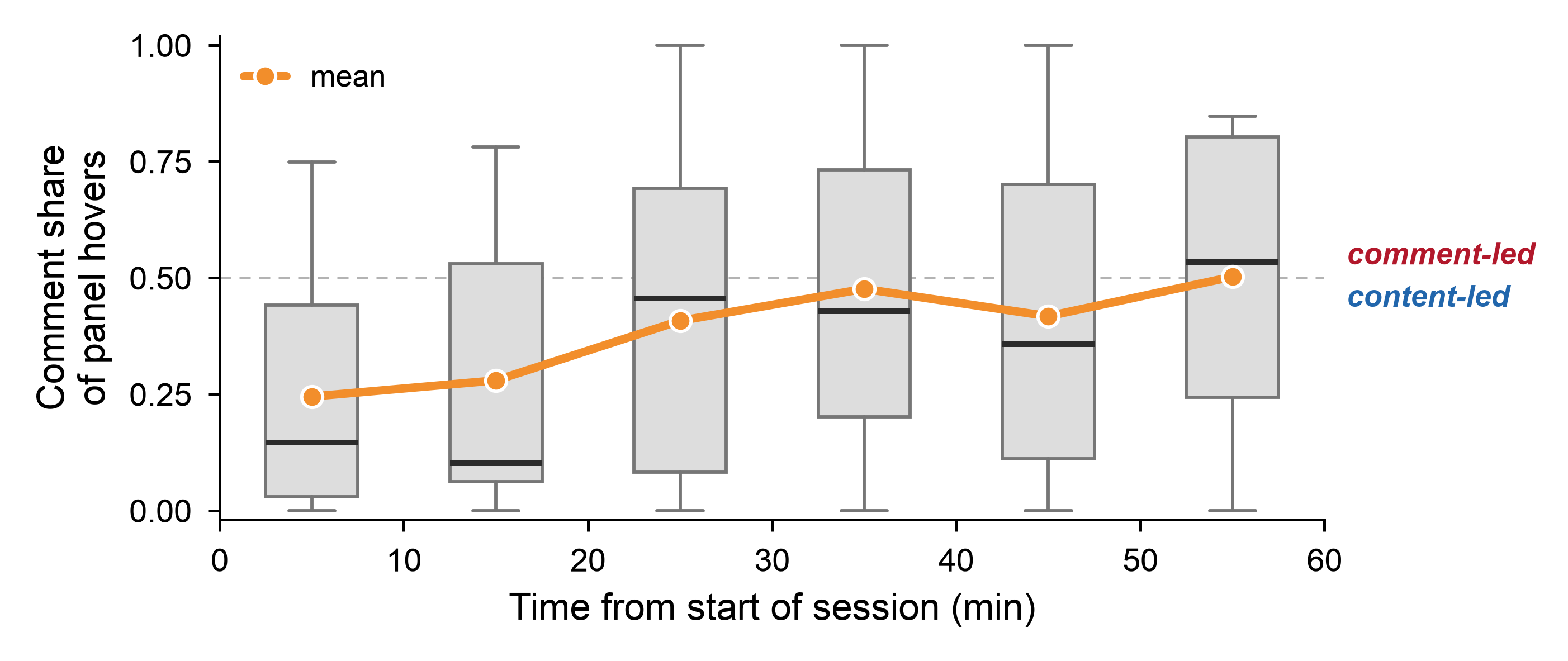}
        \caption{Comment share of panel hovers over the session in SocialWeave.}
        \label{fig:treehover}
    \end{subfigure}    
    \caption{Temporal engagement with social comments in SocialWeave}
\end{figure}

Having located consultation within the structure of a decision, we next characterize \textit{when} the comments are consulted in the process. Consultation shifted toward the comment channel as sessions progressed. The comment share of hovers within each participant increased over the session (Figure \ref{fig:treehover}). Participants hovered the comment panels for around 25\% in the first ten minutes and rose steadily over time, reaching approximate parity with the content channel by the end of the hour.

Video commitment via the comment channel followed a similar distribution. 80\% of SocialWeave participants first selected a video via comments panels after the 15-minute mark, with the aggregate share of video selection from the comment panels also rising after 15 mins (Figure \ref{fig:treeclick}). P313 explain their reason: \textit{"To decide my watching sequences, I started from the left (content cues) since it's clear, and when I saw the comment when watching videos, I found Commentary Treemap (social cues) was also helpful and clicked on those video recommendations."} The comments are more consulted once participants have some impression about the video topic, to interpret the comments.

With all the analyses about the consultation of comments, not all participants appreciated the new information channel summary. 15 participants using SocialWeave selected and fully watched at least one video chosen from the Commentary Recommend List; the remaining 5 participants either didn't select a video from the comment-related panels. Interviews reveal that this variation is associated with participants' assessments of audience comments as a source for video navigation.

Comments could feel noisy, partial, or distracting, as said by participants. Three participants who hesitated to rely on comment-derived panels raised concerns about the quality and relevance of comments. Some found comments ``\textit{off-topic}'' or overly subjective (e.g., P308), while others felt comments could overemphasize only a portion of a video: “\textit{I felt the comments only related to a part of the video, which may not represent the whole content}” (P307). A few participants preferred content cues over social cues because they perceived the latter as less structured for navigation and as less useful when presented as “\textit{bullet point styles only}” (P309). For these participants, social cues were either used sparingly or downweighted relative to content structure. P313 also emphasized that they used comment cues while remaining cautious about overtrusting them for forming an opinion.

Overall, these evidences highlight a stance-dependent usage of comments for video navigation consultation, although aggregated by computational pipeline: participants who were skeptical of comments tended to rely primarily on content structure, while those who saw value in audience comments often adopted the comment channel after initial orientation to prioritize, detour, and occasionally challenge their evolving understanding.


\section{Discussion}

Comments form a rich audience-reaction signal that can shape how users navigate videos. In this work, we explored this idea by designing SocialWeave, a comment-guided interface that presents structural overviews of both video content and community comments, and by studying how participants adopted these comment-based summaries during a navigation task.

\subsection{Social Comments for Video Navigation}

SocialWeave integrates structural social cues from comments to complement video content-based guidance. Our findings reveal that these comment-based overviews influenced how participants navigated and interpreted videos in nuanced ways.  Compared with content-based cues, social cues introduce a distinctly social dimension for navigation or attentional breaks. Participants shared that audience comments sometimes detoured their viewing flow, yet in many cases, these moments prompted them to refine their navigation strategies. These consideration didn't add additional cognitive load to users according to NASA-TLX. 
These findings suggest that comments from weakly-connected communities can foster navigation decisions, supported with computational pipelines. Notably, the value of this channel did not lie in re-presenting comments participants could have read. The video material analysis reported that most of the social signals participants consulted were effectively below the fold compared to the default presentation. The structured overview converted messages from comments that were practically inaccessible into scannable, actionable scent.

At the same time, participants’ attitudes toward using these comments were mixed, reflecting heterogeneity in how they weighted expected information gain against risk and credibility costs. They are mediated by trust in audience comments, compared to content, which was largely taken at face value by participants. This echoes earlier findings on user trust in viewer-generated summaries in livestream learning contexts \cite{lu2018streamwiki}, suggesting lower adoption of summaries created by viewers due to authenticity concerns. Trust in audience traces may modulate the extent to which participants convert comment cues into foraging actions (entering a video, detouring, or abandoning a path) and the approach they take \cite{hatamleh2023trust}.

Taken together, our findings highlight the tension for social comments in weakly-connected communities for navigation: the same signals that foster active navigation breaks and prompt reconsideration of candidate videos can also introduce uncertainty about content credibility. This tension may be especially pronounced in higher-stakes information-seeking settings, such as learning or news contexts, than in music or movie recommendation. Beyond simply \textit{surfacing more social signals} with algorithmic support, future systems may consider how to calibrate social scent from trust and attitude angles to support more human-centric navigation.

\subsection{Temporal Adoption of Comments}

From the quantitative and qualitative results, we also reveal key timings for participants to engage with social comments for navigation.  

The first uptake of social cues usually came after participants had formed an initial impression of the topic. Our findings indicate that many participants began by orienting with content cues, and only after initial browsing or early viewing did they turn to comment panels. One possible reason for this sequencing is that social cues become more interpretable once users have established a provisional frame of the topic space. With enough context, users may start judging whether comment themes are relevant and useful and deciding when to adopt the social cues.

When participants engaged with social cues in SocialWeave, their use clustered around routing options: moments of deciding which directions to watch next, whether to go deeper, or when to change direction.  From the SIFT view, social cues operated primarily as scent for patch switching rather than as overall video guidance: they helped participants re-rank options and revise exploration during transitions, which helps explain how these cues could interrupt flow yet still translate into purposeful rerouting rather than wasted exploration, without eroding the structured navigation. 
These temporal patterns suggest the potential for a staged integration of social cues into navigation systems. When coordinated views encode heterogeneous cue types, they don't need to share equal salience throughout a session. Rather than surfacing comment cues at maximum salience from the outset, interfaces could adopt progressive, adaptive visibility. Aligning the presentation of social cues with users’ readiness and decision moments may preserve the benefits of community-derived social cues, while reducing early overload and perceived noise.

\subsection{Human-centric Algorithm Support}

Our findings suggest that the challenge of integrating information scent is not only technical or related to aggregation quality. Multiple factors beyond the visualization and algorithmic recommendations impacted whether and how participants engaged with comment-derived signals: trust in community voices, evolving topic understanding, and navigational context. This is consistent with research on algorithm aversion or appreciation, where the adoption is moderated by interaction mode, framing, temporal considerations and decision stakes \cite{jussupow2024integrative}. Supporting navigation with algorithmically processed signals, then, is as much a question of human-centered design as of computational technique.

SocialWeave responds to this challenge through design choices that prioritize interpretability and source transparency. Rather than automating navigation, the computational pipeline restructures dispersed community discourse into scannable themes (\textbf{R1}); the two-layer visualization preserves links back to source comments and videos, enabling users to drill into details (\textbf{R2}). Our study findings suggest these transparency affordances are helpful: participants consulted comments most purposefully at moments when they could ground those cues in the videos and in their understanding.

However, transparency supports interpretation but can not guarantee accuracy. Comments can be topically relevant yet misleading, and presenting them within an organized, system-curated overview risks lending such content unwarranted authority. Our findings suggest a mitigation already latent in the design: participants spontaneously used provenance links to cross-check cues when comments diverged from a video's framing, turning the same links that support navigation into a verification pathway. Future systems could make this pathway more explicit, for example by pairing comments with verified source \cite{bode2015related} or by surfacing disagreement among commenters rather than smoothing it away \cite{kriplean2012supporting}.

Advances in LLMs is expected to sharpen both sides of this picture. Recent advances in more accessible and powerful LLMs have largely empowered recommendation and navigation systems. They may encode more signals into the consideration: richer signals, more media types, and more flexible representations \cite{kim2023surch, wang2024surgment, dimarco2023llm}.
Yet, generative AI systems increasingly mediate not only access to information but also its presentation and interaction, reshaping and defining the relevance and visibility in the process \cite{yang2025videomix}, the complexity of algorithmic results adoption and usage strategy also becomes prominent. 
Better aggregation alone will therefore not settle how content and signals are adopted. The question our findings pose for AI-augmented search is how to encode and present social and contextual signals so that they remain not only well-aggregated but interpretable, verifiable, and attuned to the user's navigational context.

\subsection{Limitation and Future Work}

When designing SocialWeave, we chose treemaps to convey the structure of audience comments, supporting an “overview first, then details” exploration workflow. 
One participant (P309) perceived Commentary Treemap as under-structured, suggesting that the visual abstraction for large scale of short user-generated content still have opportunities.  Future work could examine alternative representations and interaction techniques to improve the experiences with comment summary.

Our lab study has some limitations. We selected a YouTube playlist as an alternative to SocialWeave and Content-centric Prototype, while there are other natural options, such as learning on YouTube directly. We chose it to control the video pool participants may access; it turned out to lead to similar concept coverage, given the salience of content on general-purpose video platforms. Future work may explore other real-world alternatives and integrate with emerging AI-driven information-seeking processes. 

Inspired by our findings on enabling appropriate community-derived cues with temporal consideration, we plan to further personalize the video navigation and provide adaptive recommendations in real time. For instance, progressive overview guidance with varying levels of detail could cater to navigation progress and improve their navigation experiences with sufficient but not overwhelming cues. Other user feedback and navigation traces could also be utilized for video navigation support, such as previous users' paths \cite{mu2019moocad}. We also plan to leverage human-in-the-loop Retrieval-Augmented Generation LLMs with domain knowledge to simulate high-quality comments, thereby improving the quality of comment-based cues to further evaluate user adoption \cite{liu2018conceptscape}. 

From the domain of the resource, this study focuses on helping people navigate generic multi-media platforms, like YouTube, as avenues for informative, factual content. In contrast to YouTube \cite{shoufan2022youtube}, platforms like MOOCs and TikTok differ in content scope, community norms, trust relationships, and interaction dynamics, which may substantially reshape how users interpret and respond to these social cues. Another line of future work is planned to examine how these social variables mediate decision-making strategies and outcomes on other interaction mechanisms with GenAI systems, both in video exploration and in other knowledge tasks.

\section{Conclusion}

Cross-video navigation on general-purpose platforms increasingly exposes users to community comments, yet most of them are not accessible for navigation. We designed SocialWeave around the idea that comment content, when computationally distilled into structural thematic overviews, can meaningfully complement content-based navigation guidance. 
A controlled lab study with 60 participants showed that adding comment-derived overview and recommendations opens a parallel consultation channel that prompted rerouting at decision moments. We also revealed how these cues were adopted in practice selectively at routing moments, and contingent on users' trust in community voices and information sources. These findings characterize algorithm-mediated community comments as trust-dependent, phase-sensitive navigational resources, and point toward AI-augmented search interfaces that surface social signals adaptively, calibrated to users' readiness and decision contexts.


\bibliographystyle{ACM-Reference-Format}
\bibliography{cgp_bib}

\appendix

\section{Computational pipeline design details}\label{appendix: comptutation}

\textbf{Relation Mapping:} Four aspects are used to discern prerequisites between pairs of concepts $(a, b)$ using rule-based approaches. It's built upon methodologies introduced by Pan et al. \cite{pan2017prerequisite} and modified in \cite{tang2021conceptguide}.

\textit{- Semantic similarity (Ss).}  it's the average of three metrics: cosine similarity of word embeddings, the co-occurrence ratio of a and b across the video corpus (local closeness), and a Wikipedia-based score assigned when b is referenced in a's Wikipedia entry (global closeness from external resources).


\textit{- Video reference prerequisite (Vrp) and Wikipedia reference prerequisite (Wrp).} 
Both assess prerequisite direction through the relative positions of a and b in videos and Wikipedia pages, respectively. A non-zero value infers a prerequisite between a and b.

\begin{equation}
\begin{split}
Vrp(a,b) = GVrw(b,a)-GVrw(a,b),\\
GVrw(a,b) = \frac{\sum_{i=1}^{M} Vrw(a_{i},b) \cdot w(a_{i},b)}{\sum_{i=1}^{M} w(a_{i},b)}, \\
Vrw(a,b) = \frac{\sum_{v\in{V}} f(a,v) \cdot r(b,v)}{\sum_{v\in{V}} f(a,v)}, \\
\end{split}
\label{eqn:Vrp}
\end{equation}

\textit{- Complexity level distance (Cld).}  
This metric captures another facet of prerequisite relationships by examining the frequency and temporal distribution of concepts across videos. \( Cld(a, b) = avc(a) - avc(b) \), where \( avc(a) \) represents the proportion of videos in \( V \) that refer to concept $a$.


The measures are synthesized into a composite score $P(a, b)$ through weighted summation, normalized into [0, 1]. A positive value signifies that a is a prerequisite of b, and vice versa; pairs exceeding a predetermined threshold are rendered as directed links in the Content Diagram.

\textbf{Comment Classifier Training Details}
We customize a classifier on ERNIE, a knowledge-enhanced transformer-based language model with strong fine-tuning and zero-shot performance on natural language understanding tasks \cite{sun2021ernie}. all ERNIE layers are frozen and an additional layer is added over the output. To create training data, two independent coders manually collected and coded a small collection of YouTube comments with labels such as "greetings," "questions," and "opinions," following the classification scheme in \cite{madden2013classification}, and regrouped them into binary labels with respect to the video topic — e.g., "greetings" and "spam" merged as irrelevant; "summary" and "opinion" merged as relevant. The classifier was fine-tuned few-shot on this labeled dataset.



\end{document}